\documentstyle[aps,multicol,fleqn,psfig]{revtex}

\newcommand{\beq}{\begin{equation}}
\newcommand{\eeq}{\end{equation}}
\newcommand{\beqa}{\begin{eqnarray}}
\newcommand{\eeqa}{\end{eqnarray}}
\newcommand{\ket} [1] {\vert #1 \rangle}
\newcommand{\bra} [1] {\langle #1 \vert} 
\newcommand{\braket}[2]{\langle #1 | #2 \rangle}

\newcommand{\mean}[1]{\langle #1 \rangle}

\begin{document}
\title{Phase-Conjugated Inputs Quantum Cloning Machines}
\author{N. J. Cerf$^{1,2}$ and S. Iblisdir$^1$}
\address{$^1$ Ecole Polytechnique, CP 165, Universit\'e Libre de Bruxelles,
B-1050 Bruxelles, Belgium\\
$^2$ Jet Propulsion Laboratory,
California Institute of Technology, Pasadena, CA 91109\\}
\date{February 2001}

\draft

\maketitle

\begin{abstract}
A quantum cloning machine is introduced that yields $M$ identical 
optimal clones from $N$ replicas of a coherent state and $N'$ replicas of its
phase conjugate. It also optimally produces $M'=M+N'-N$
phase-conjugated clones at no cost. 
For well chosen input asymmetries $N'/(N+N')$, this machine is shown 
to provide better cloning fidelities than the standard 
$(N+N') \to M$ cloner. The special cases of the optimal 
balanced cloner ($N=N'$) and the optimal measurement ($M=\infty$) 
are investigated.
\end{abstract}


\begin{multicols}{2}
\narrowtext

The concept of cloning plays a central role in quantum information
theory. It is for example crucial in quantum cryptography, as the
optimal duplication of a quantum state directly determines
the security of a cryptosystem\cite{fuch97,cerf00:qc}, or 
in quantum estimation theory, as cloning provides a constructive way 
to achieve a generalised measurement (POVM) \cite{buze98}. The optimal
$N$-to-$M$ cloning transformations, which produce $M$ clones from
$N$ originals, have been found for both quantum bits\cite{gisi97}
and continuous variables\cite{brau00,fiur00}. Interestingly,
producing infinitely many clones from $N$ identical replicas 
of a quantum state is equivalent to performing an optimal measurement,
which reflects the existence of a close link between cloning 
and measurement theories. In the context of measurement, recent work
has revealed that pairs of antiparallel qubits are intrinsically more
informative than pairs of parallel qubits \cite{gisi00}, a result that
has later been extended to continuous variables: more information can be
encoded in a pair of phase-conjugated coherent states
$\ket{\psi}\ket{\psi^*}$ than in two identical replicas
$\ket{\psi}\ket{\psi}$ \cite{cerf00:pc}. This property suggests
that cloning machines with antiparallel input qubits (or phase-conjugated 
input modes) might yield better fidelities than standard $N \to M$ 
cloning machines, thereby opening a new avenue 
in the investigation of quantum cloning.

In this Letter, we will focus on quantum information carried by
continuous variables, and seek for a cloning transformation that, taking 
as inputs $N$ replicas of a coherent state $\ket{\psi}$ 
and $N'$ replicas of its complex conjugate $\ket{\psi^*}$,
produces $M$ optimal clones of $\ket{\psi}$. 
The resulting concept of phase-conjugated inputs (PCI) cloning machines
will turn out to be closely connected to that
of the amplification of light, just as 
for standard cloning \cite{mand83,simo00,brau00,fiur00}.  
As a matter of fact, it can be decomposed as a
sequence of beam-splitters, a single non-linear medium, and another
sequence of beam-splitters, which is consistent with the Bloch-Messiah
reduction theorem \cite{brau99}. We will start by deriving the optimal 
canonical transformation that acts on two modes in a coherent state
with respective mean values $\alpha \psi$ and $\beta \psi^*$ (where
$\alpha, \beta$ are real while $\psi$ is a c-number), and generates
a mode whose mean value is $\gamma \psi$, where $\gamma$ is real.
Remarkably, this transformation will be shown to have 
a structure similar to that of a conventional phase-insensitive
phase-preserving amplifier as defined in \cite{cave82}, where both
the signal and idle modes are used as inputs. After having
derived this transformation, we will apply it to the case of integer
$\alpha^2$, $\beta^2$, and $\gamma^2$, and see how it can be
supplemented with beam-splitters to provide
a PCI cloning machine for continuous variables. 
This machine will be shown to produce $M'=M+N'-N$ additional
phase-conjugated clones (or anticlones). The quality of the
clones and anticlones will be discussed in the case of a balanced cloner
($N=N'$), as well as for arbitrary input asymmetries $N'/(N+N')$. The
related question of the optimal measurement ($M=\infty$) will also
be treated. To our knowledge, the PCI cloner is the first example of a 
new quantum information-theoretic process for continuous variables,
for which no discrete-variable analogue has been found yet.

Let $\{ a_i \}$ and $\{ b_i \}$ ($i=1\ldots 3$) respectively denote
the input and output modes annihilation operators of the cloning
transformation. The indices $i=1,2$ respectively refer to the
input and phase-conjugated input modes, while $i=3$ refers to an
auxillary mode. In full generality, we are seeking for a linear
canonical transformation
\beq
b_i=M_{ij} a_j + L_{ij} a_j^{\dagger} \qquad (i,j=1 \ldots 3),
\eeq
that meets the three following requirements (the sum over repeated 
indices being implicit). Firstly, starting with modes $a_1$ and $a_2$ 
with mean values $\mean{a_1}=\alpha \psi$ and $\mean{a_2}=\beta \psi^*$, we
require $\mean{b_1}=\gamma \psi$. We will only consider the
case $|\gamma| \geq |\alpha|$, since, otherwise, the problem becomes
trivial: one would just have to attenuate the input coherent state
$\ket{\alpha \psi}$ with an unbalanced beam-splitter, yielding a
coherent state of amplitude $\gamma \psi$. To simplify the
problem, we may assume that $\beta=1$, which amounts to susbtitute
$\psi$ for $\beta \psi$. Then, we have:
\beqa\label{eq:mean}
\alpha M_{11}+L_{12} &=& \gamma, \nonumber \\
M_{12}+\alpha L_{11} &=& 0 .
\eeqa
Secondly, this transformation must obey the commutation rules
$[b_i,b_k]=0$ and $[b_i,b_k^{\dagger}]=\delta_{ik}$ ($\hbar=1$), that is
\beqa\label{eq:cr}
M_{ij}L_{kj}-L_{ij}M_{kj} &=& 0, \nonumber \\
M_{ij}M_{kj}^*-L_{ij}L_{kj}^* &=& \delta_{ik}.
\eeqa
Thirdly, the noise of the output mode $b_1$ of this transformation
should be minimum.

Before sketching our calculation, let us note that a further
simplification comes from the fact the the annihilation operators are
defined up to an arbitrary phase, so that a transformation $a_i\to
e^{i\mu_i} a_i$ and $b_1\to e^{i \nu} b_1$ allows us to take $M_{1j}$
and $L_{1j}$ as real and positive. Since we focus on phase-insensitive
transformation, minimizing the noise amounts to minimizing the sole
quantity $(\Delta b_1)^2=\frac{1}{2}\mean{b_1
b_1^{\dagger}+b_1^{\dagger} b_1}-\mean{b_1}\mean{b_1^{\dagger}}$
\cite{cave82}.  Thus, using the fact that $(\Delta a_i)^2=1/2$ 
for a mode $a_i$ in a coherent state, we need to minimize
\beq\label{eq:var}
(\Delta b_1)^2=\frac{1}{2}(M_{1j}M_{1j}+L_{1j}L_{1j}),
\eeq
under the constraints Eqs.~(\ref{eq:mean}) and (\ref{eq:cr}). Rather
than solving this full problem, we use here a common trick in
constrained extremization problems that consists in solving a simpler
problem with weaker constraints (bearing in mind that taking weaker
constraints can only yield better solutions) and then checking that
the solution of this simpler problem is one of the full
problem. Specifically, we minimize $(\Delta b_1)^2$ taking into
account the only condition $M_{1j}M_{1j}-L_{1j}L_{1j}=1$. Taking
Eq.~(\ref{eq:mean}) into account and introducing a Lagrange
multiplier $\lambda$, we minimize the quantity
$M_{11}^2+(\gamma-\alpha
M_{11})^2+(1+\alpha^2)L_{11}^2+M_{13}^2+L_{13}^2+\lambda
\big(M_{11}^2-(\gamma-\alpha
M_{11})^2-(1-\alpha^2)L_{11}^2+M_{13}^2-L_{13}^2-1 \big)$, with
respect to $M_{11},L_{11},M_{13}$ and $L_{13}$. Some algebra shows
that this problem admits only one solution
$M_{13}=L_{13}=L_{11}=M_{12}=0$, that is, the auxillary mode is
unnecessary.  The optimal transformation has then the same structure
as that of a phase-insensitive amplifier of gain $G$. Restoring
$\beta$, we get
\beqa\label{amplifier}
b_1 &=& \sqrt{G} a_1 +\sqrt{G-1} a_2^{\dagger}, \nonumber \\
b_2 &=& \sqrt{G-1} a_1^{\dagger}+\sqrt{G} a_2,
\eeqa
with
\beq\label{gain}
\sqrt{G}=\frac{-\alpha \gamma+\beta \sqrt{\gamma^2-\alpha^2+\beta^2}}
{\beta^2-\alpha^2},
\eeq
It can easily be checked that, for $\beta=0$ (or $\alpha=0$),
Eq.~(\ref{amplifier}) reduces to a phase-insensitive
phase-preserving (or phase-conjugating) amplifier as defined in
\cite{cave82}, and can be used to carry out the $N \to M$ cloning 
(or phase-conjugating) transformation described in \cite{brau00} 
(or \cite{cerf00:pc}).

\begin{figure}
\caption{PCI cloner that produces $M$ clones and $M'$ anticlones
from $N$ replicas of $\ket{\psi}$ and $N'$ replicas of $\ket{\psi^*}$.
Modes are concentrated and distribitued by Discrete Fourier Transform (DFT). PCIA stands for a phase-conjugated inputs amplifier.} 
\vskip 0.25cm
\centerline{\psfig{file=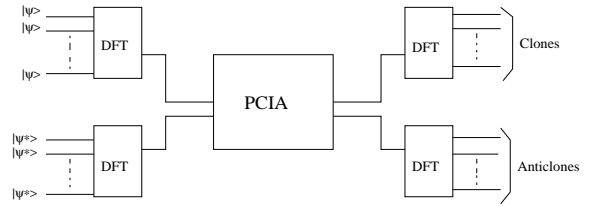,width=3.0in,angle=-90}}
\label{fig:pcic}
\vskip -0.25cm
\end{figure}

Let us now turn to the special case where $\alpha^2, \beta^2$ and
$\gamma^2$ are integers (which we will denote respectively as
$N,N',M$). The transformation Eq.~(\ref{amplifier}) can be used as the
central element of a PCI cloning machine, which is covariant for
translations and rotations in phase space (see Fig.~\ref{fig:pcic}). 
Indeed, the following procedure can be used to produce $M$ optimal clones 
of a coherent state $\ket{\psi}$ from 
$\ket{\psi}^{\otimes N} \ket{\psi^*}^{\otimes N'}$:

(i) Concentrate the $N$ replicas of $\ket{\psi}$ stored in the $N$
modes $\{c_l\}$ ($l=0 \ldots N-1$) into a single mode $a_1$, this
results in a coherent state of amplitude $\sqrt{N}\, \psi$. This
operation can be performed with a network of beam-splitters
achieving a $N$-mode Discrete Fourier Transform (DFT)\cite{brau00}, 
which yields the mode
\beq
a_1=\frac{1}{\sqrt{N}}\sum_{l=0}^{N-1} c_l,
\eeq
and $N-1$ vacuum modes. Similarly, concentrate the $N'$ replicas of
$\ket{\psi^*}$ stored in the $N'$ modes $\{d_l \}$ $(l=0 \ldots N'-1)$,
into a single mode $a_2$ in a coherent state of amplitude
$\sqrt{N'}\, \psi^*$, with the help of a $N'$-mode DFT:
\beq
a_2=\frac{1}{\sqrt{N}}\sum_{l=0}^{N'-1} d_l.
\eeq

(ii) Process the modes $a_1$ and $a_2$ into a ``phase-conjugated
inputs'' amplifier (PCIA), resulting in modes $b_1$
and $b_2$ as defined in Eqs.~(\ref{amplifier}) and (\ref{gain}).

(iii) Distribute the output $b_1$ into $M$ clones $\{c'_l\}$ $(l=0
\ldots M-1)$ with a $M$-mode DFT:
\beq\label{eq:dft}
c'_l=\frac{1}{\sqrt{M}}(b_1+e^{i \pi kl/M} v_k),
\eeq
where $\{v_k \}$ $(k=1 \ldots M-1)$ denote $M-1$ vacuum modes. 
It is readily verified that this procedure yields 
$M$ clones of $\ket{\psi}$.

Interestingly, the amplitude $b_2$ of the other output of the PCIA
has a mean value $\sqrt{M'}\psi^*$, with
\beq\label{eq:balance}
N-N'=M-M'.
\eeq 
Therfore, it can be used to produce $M'$ phase-conjugated clones 
(or anticlones) of $\ket{\psi}$,
$\{d'_l\}$ $(l=0 \ldots M'-1)$, using a $M'$-mode DFT:
\beq\label{eq:dft2}
d'_l=\frac{1}{\sqrt{M'}}(b_2+e^{i \pi kl/M} w_k)
\eeq
where $\{w_k\}$ $(k=1 \ldots M'-1)$ denote $M'-1$ vacuum modes. 
Clearly, this procedure is optimal to produce $M$ clones since its
central element, the PCIA, is optimal, and the beam-splitters are
passive elements. In addition, the $M'$ anticlones that are produced
at no cost are also optimal. Indeed, our transformation produces 
$M$ optimal clones and $M \geq N$, and is symmetric with respect to the
interchange of labels $1$ and $2$. So, if our initial problem 
was to produce $M'$ optimal anticlones with $M' \geq N'$, 
we would find the same solution. Noting that that $M \geq N
\iff M' \geq N'$  [see Eq.~(\ref{eq:balance})], it therefore is clear
that our transformation yields both optimal clones
and optimal anticlones. Furthermore, since the PCIA
is linear and phase-insensitive, the resulting PCI cloner is covariant with
respect to translations and rotations of the state to be copied: all
coherent states are copied equally well, and the cloning-induced
noise is the same for all quadrature components.

Using Eqs. (\ref{amplifier})-(\ref{eq:dft}) and (\ref{eq:dft2}), 
the noise of the clones and anticlones can be written as
\beq\label{eq:clonoise}
(\Delta {c'_l})^2=\frac{1}{2}+\frac{G-1}{M} ,  \quad
(\Delta {d'_l})^2=\frac{1}{2}+\frac{G-1}{M'} ,
\eeq 
where the gain can be reexpressed as a function of the number
of inputs and outputs,
\beq
\sqrt{G}=\frac{\sqrt{N'M'}-\sqrt{NM}}{N'-N}
\eeq
As expected, the variance of the output clones exceeds $1/2$, implying
that the clones are not exactly in the coherent state
$\ket{\psi}$. Instead, they suffer from a thermal noise with a mean
number of photons given by $\mean{n_{th}}=(G-1)/M$. In other words,
their $P$-function \cite{scul97} is a Gaussian distribution
\beq\label{eq:pfunc}
P(\xi,\xi^*)=\frac{1}{\pi \mean{n_{th}}} \; e^{-|\xi-\psi|^2/\mean{n_{th}}}.
\eeq
rather than a Dirac distribution $P(\xi,\xi^*)=\delta^{(2)}(\xi-\psi)$.

Consider now the balanced case ($N=N'$, $M=M'$), for which simple
analytical expressions of the noise variances
can be obtained. Taking the limit $\alpha \to \beta$ in
Eq.~(\ref{gain}) and replacing $\alpha^2$ by $N$ and $\gamma^2$ by
$M$ yields $G=(M+N)^2 / 4MN$, so that the error variances 
of the clones and anticlones are
\beq\label{eq:pcinoise}
(\Delta {c'_l})^2=(\Delta {d'_l})^2= \frac{1}{2}+\frac{(M-N)^2}{4M^2N}.
\eeq 
Note that this balanced cloner is optimal among all
PCI cloners in the sense that it minimizes $\mean{n_{th}}$
for fixed $N+N'$ and $M+M'$.
It is convenient to characterize the quality of cloning in terms of
the fidelity $f_{{N\choose N}\to M}=\bra{\psi} \rho_c
\ket{\psi}/|\braket{\psi}{\psi}|^2$ where $\rho_c$ denotes the state
of the clones. Using Eq.~(\ref{eq:pfunc}), we get
\beq\label{eq:balfid}
f_{{N\choose N}\to M}=\frac{1}{1+\mean{n_{th}}}=\frac{4M^2 N}{4M^2 N+(M-N)^2}.
\eeq
Let us now compare the production of $M$ clones from $N$ replicas and $N$
antireplicas to the production of $M$ clones from $2N$ identical
replicas. The variance and fidelity of the clones $k_i$ obtained by standard
cloning are given by\cite{cerf00:oc} 
\beq \label{eq:stdnoise}
(\Delta {k'_i})^2=\frac{1}{2}+\left(\frac{1}{2N}-\frac{1}{M}\right), 
\eeq
and
\beq 
f_{2N\to M}=\frac{2MN}{2MN+M-2N}.  
\eeq 
Of course, in the trivial case where $M=2N$, standard cloning can be
achieved perfectly, while the balanced PCI cloner yields an additional
variance $\mean{n_{th}}=1/(16N)$. However, whenever $M \geq 2N+1$,
the $({}^{N}_{N})\to M $ balanced cloner 
always yields a lower variance (hence a higher fidelity)
than the $2N \to M$ cloning machine. The balanced PCI cloner is also
better for the anticlones: more anticlones are produced at no cost,
and they have a better fidelity. Indeed, a
standard $2N\to M$ cloning machine produces $M-2N$ anticlones of
fidelity $2N/2N+1$, which actually is the fidelity of an optimal measurement 
of $2N$ replicas of $\ket{\psi}$. In contrast, a PCI cloner
produces $M$ anticlones with a higher fidelity, as given by
Eq.~(\ref{eq:balfid}). In particular, for a measurement
($M=\infty$), we see from Eqs.~(\ref{eq:pcinoise}) and
(\ref{eq:stdnoise}) that the additional noise induced by a PCI
cloner is $1/4N$, that is, {\em one half} of the noise induced by a standard
$2N \to \infty$ cloner ($1/2N$). This reflects that \emph{more
information is encoded in $N$ pairs of phase-conjugated replicas of a
coherent state than in $2N$ identical replicas}, a result which was
proven for $N=1$ in \cite{cerf00:pc}. More generally, in the
unbalanced case ($N\ne N'$), it can be shown that
the optimal measurement results in a noise 
that is equal to that obtained by measuring
$(\sqrt{N}+\sqrt{N'})^2$ identical replicas of the input, 
in the absence of phase-conjugated inputs.

We have shown that the balanced PCI cloner results in
better cloning fidelities than a standard cloner. More generally,
we may ask the following question: \emph{If we want to produce $M$
clones of a coherent state $\ket{\psi}$ from a fixed total number $n$
of input modes, $N$ of which being in the coherent state $\ket{\psi}$
and $N'$ of which being in the phase-conjugated state $\ket{\psi^*}$,
what is the asymmetry $a=N'/n$ that
minimizes the error variances of the clones?} 

\begin{figure}
\caption{Cloning-induced noise standard deviation $\sqrt{\mean{n_{th}}}$ 
as a function of the fraction of the asymmetry $a=N'/n$,
for $n=8$ and several values of $M/n$.}
\vskip 0.25cm
\centerline{\psfig{file=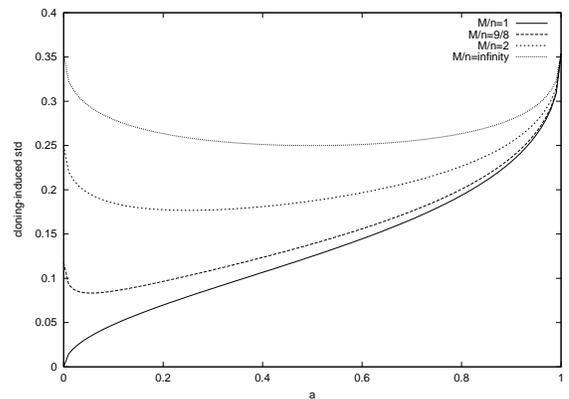,width=3.0in,angle=-90}}
\label{fig:gder}
\vskip -0.25cm
\end{figure}

From Eq.~(\ref{amplifier}), we see that for fixed values 
of the total number of inputs $n$ and number of outputs $M$, 
the gain $G$ (and thus the thermal noise of the clones
$\mean{n_{th}}$) depends only on $a$, and varies as
\beq\label{eq:gr}
G(a)=\left(\frac{
\sqrt{a} \sqrt{\frac{M}{n}+(2a-1)}
-\sqrt{\frac{M}{n}} \sqrt{1-a}
}{2r-a}
\right)^2
\eeq 
In Fig.~(\ref{fig:gder}), we have plotted $\sqrt{\mean{n_{th}}}$ as a
function of $a$ for $n=8$ and different values of $M\geq n$. (Of course,
only rational values of $a$ are relevant here, but the whole curve has been
plotted for simplicity.) In the trivial case where $M=n=8$,
the minimum additional variance is of course zero, and is obtained for
$a=0$. The cloning transformation is then just the identity.
However, when $M \geq n+1$ , using
phase-conjugated input modes yields lower variances than standard cloning
if $a$ is correctly chosen (the lowest variance is attained for $a
\neq 0$).  Remarkably, the value of $a$ achieving the minimum variance 
is not equal to $1/2$ for finite $M$, that is the optimal input partition
contains more replicas than antireplicas. In the limit of large $M$, 
however, the number of antireplicas achieving
the lowest variances tends to $n/2$, and the curve $G(a)$
tends to a symmetric curve around $a=1/2$. This symmetry is
not surprising, since $M =\infty$ corresponds to a measurement
\cite{gisi97,cerf00:oc} and we expect that measuring the value of
$\psi$ from $N$ replicas of $\ket{\psi}$ and $N'$ replicas of
$\ket{\psi^*}$ is equivalent to starting from $N'$ replicas of
$\ket{\psi}$ and $N$ replicas of $\ket{\psi^*}$. So, we have found that
the optimal measurement is achieved with balanced inputs ($N=N'$).
Finally, in the case where $a=1$, 
the transformation consists in producing $M$ clones of
$\ket{\psi}$ from $n$ replicas of $\ket{\psi^*}$. This is just
phase-conjugation, for which we know that the best strategy is to
perform a measurement\cite{cerf00:pc}. 
The additional variance is therefore given by $1/n$,
which does not depend on $M$. This explains why 
the curves converge all to the same point at $a=1$.

In summary, we have derived a continuous-variable cloning
transformation using phase-conjugated inputs. This transformation
has been shown to be decomposable in a sequence of beam-splitters, a
central amplification stage, and another sequence of beam-splitters. A
possible way to implement this central stage would be to use four-wave 
mixing. Two weak fields entering the $\chi^{(3)}$ medium
would then play the role of the phase-conjugated inputs, and energy
would be brought to the system by two external modes in a large
coherent state (see \cite{scul97} for details). We have shown evidence
that PCI cloning transformations outperform
standard cloning transformations (taking only identical inputs) 
if the goal is to produce clones and anticlones of a state or to get
knowledge about a state through measurement. The special case of
the balanced cloner, which produces $M$ pairs of phase-conjugated clones
from $N$ pairs of phase-conjugated replicas, has been analyzed
and was shown to be optimal. 
As far as we know, no qubit analogue of our transformation has been
proposed yet, though it is very plausible that a cloning machine
that produces $M$ clones from $N$ qubits in an arbitrary state
and $N'$ qubits in its orthogonal state can be defined. 
Another possible extension of
this work would be to study the case where the number of anticlones
is a free parameter (in the PCI cloner derived here, it
is constrained by $N$, $N'$, and $M$). 
Also, an interesting generalization would be to investigate fully
asymmetric PCI cloning transformations, that is, transformations
whose output clones have all different fidelities. 
Finally, our work raises the question of why can a better
cloning be achieved when phase-conjugated inputs are available, 
rather than identical inputs.
It is known that, in standard cloning, the spontaneous
emission occuring during the amplification stage is the physical
mechanism that hinders perfect cloning. Thus,
an open question left by our work is to understand why things happen
as if comparatively less spontaneous emission occured
in a PCI cloning machine.

N. J. C. is grateful to S. L. Braunstein and M. Hillery for very useful
discussions at the Budmerice workshop on Quantum Information in October 2000, 
which have led to the present paper. 
S. I. acknowledges support from the Belgian FRIA foundation.

\end{multicols}


\begin{references}

\bibitem{fuch97} C.A. Fuchs, N. Gisin, R. Griffiths, C.S. Niu and A. Peres, Phys. Rev. A {\bf 56}, 1163 (1997).

\bibitem{cerf00:qc} N.J. Cerf, M. L\'evy, and G. Van Assche, LANL e-print quant-ph/0008058.

\bibitem{buze98} V. Bu\v zek and M. Hillery, LANL e-print quant-ph/9801009.

\bibitem{gisi97} N. Gisin and S. Massar, Phys. Rev. Lett. {\bf 79}, 2153 (1997); R. F. Werner, Phys. Rev. A {\bf 58}, 1827 (1998).

\bibitem{brau00} S. L. Braunstein, N. J. Cerf, S. Iblisdir, P. van Loock and S. Massar, LANL e-print quant-ph/0012046.

\bibitem{fiur00} J. Fiur\'a\u sek, LANL e-print quant-ph/0012048.


\bibitem{gisi00} N. Gisin and S. Popescu, Phys. Rev. Lett. {\bf 83}, 
432 (1999); V. Bu\v zek, M. Hillery and R.F. Werner, Phys. Rev. A {\bf 60}, 2626 (R).

\bibitem{cerf00:pc} N. J. Cerf and S. Iblisdir, LANL e-print quant-ph/0012020.

\bibitem{mand83} L. Mandel, Nature (London) {\bf 304}, 188 (1983).

\bibitem{simo00} C. Simon, G. Weihs and A. Zeilinger, Phys. Rev. Lett.{\bf 84}, 2993 (2000). 


\bibitem{brau99} S. L. Braunstein, LANL e-print quant-ph/9904002. 

\bibitem{cave82} C. M. Caves, Phys. Rev. D {\bf 26}, 1817 (1982).

\bibitem{scul97} M. O. Scully and M. S. Zubairy, \emph{Quantum Optics}, Cambridge University Press (1997).

\bibitem{cerf00:oc} N. J. Cerf and S. Iblisdir, Phys. Rev. A {\bf 62}, 
040301(R) (2000).


\end{references}
\end{document}